\documentclass[superscriptaddress,twocolumn]{revtex4-1}
\usepackage{amsmath}
\usepackage{extarrows}
\usepackage[ruled]{algorithm2e}
\usepackage{graphicx}
\usepackage{dcolumn}
\usepackage{bm}
\usepackage[all]{xy}
\usepackage{indentfirst}
\usepackage{amsmath}
\usepackage{multirow}
\usepackage{mathrsfs}
\usepackage{bbm}
\usepackage{euscript}
\usepackage{amssymb}
\usepackage{extarrows}
\usepackage{bbm}
\usepackage[ruled]{algorithm2e}
\usepackage{graphicx}
\usepackage{epstopdf}
\usepackage{dcolumn}
\usepackage{bm}
\usepackage[all]{xy}
\usepackage{indentfirst}
\usepackage{amsmath}
\usepackage{multirow}
\usepackage{mathrsfs}
\usepackage{euscript}
\usepackage{amssymb}
\usepackage{appendix}

\newtheorem{theorem}{Theorem}

\newtheorem{prop}{Proposition}

\begin{document}

\title{Classifying multiparticle entanglement with passive state energies}

\author{Xue Yang}
\affiliation{School of Information Science and Technology, Southwest Jiaotong University, Chengdu 610031, China}
\affiliation{School of Computer and Network Security, Chengdu University of Technology, Chengdu 610059, China}
\author{Yan-Han Yang}
\affiliation{School of Information Science and Technology, Southwest Jiaotong University, Chengdu 610031, China}
\author{Xin-Zhu Liu}
\affiliation{School of Information Science and Technology, Southwest Jiaotong University, Chengdu 610031, China}
\author{Shao-Ming Fei}
\affiliation{School of Mathematical Sciences, Capital Normal University, Beijing 100048, China}
\affiliation{Max-Planck-Institute for Mathematics in the Sciences, 04103 Leipzig, Germany}
\author{Ming-Xing Luo}
\email{mxluo@swjtu.edu.cn}
\affiliation{School of Information Science and Technology, Southwest Jiaotong University, Chengdu 610031, China}

\begin{abstract}
Thermodynamics as a fundamental branch of physics examines the relationships between heat, work, and energy.  The maximum energy extraction can be characterized by using the passive states that has no extracted energy under any cyclic unitary process. In this paper, we focus on the concept of marginal passive state energy and derive polygon inequalities for multi-qubit entangled pure states. We show that the marginal passive state energies collectively form a convex polytope for each class of quantum states that are equivalent under SLOCC. We finally introduce multipartite passive state energy criteria to classify multipartite  entanglement under SLOCC. The present result provides a thermodynamic method to witness multipartite entanglement.

\end{abstract}

\maketitle

\section{Introduction}

Thermodynamics is a fundamental branch of physics that focuses on the relationships between heat, work, and energy in physical systems. A notable area of interest within thermodynamics is the extraction of energy from isolated quantum systems via cyclic Hamiltonian processes \cite{Allahverdyan2004,Viguie2005,Skrzypczyk2014,Perarnau2015,Vin(2016),Ciampini2017,Francica2017,Andolina2019,Monsel2020,Opatrny2021}. The maximum energy extraction requires to identify quantum states that resist energy extraction. The so-called passive state was introduced in Ref.\cite{Pusz1978}. Specially, when considering a state  with bare Hamiltonian, the basic question is how can the average energy be extracted  through a unitary transformation on the system. If the state remains passive even in the asymptotic limit, it is categorized as completely passive or thermal. The mechanism behind how passive states can yield energy from multiple copies under reversible unitary operations remains unclear. Ongoing research is focusing on energy extraction from non-equilibrium quantum states \cite{Salvia2020,Sparaciari(2017),Perarnau2015a,Alhambra2019}, the information theory framework in quantum thermodynamics \cite{Bera2019,Uzdin2018,Brown2016}, interpreting passive state energy as an entanglement measure \cite{Alimuddin2020}, the pursuit of achieving Gibbs states through geometric methods \cite{Koukoulekidis2021}, and the analysis of quantum battery capacity \cite{Yang2023}.

\begin{figure}
\begin{center}
\resizebox{230pt}{140pt}{\includegraphics{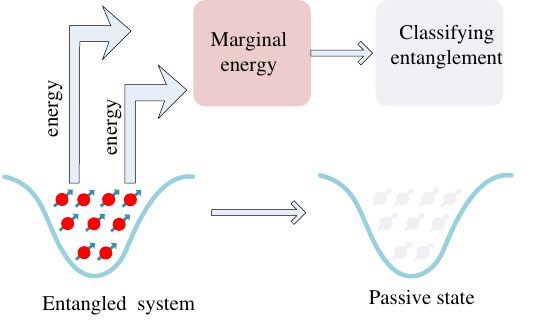}}
\end{center}
\caption{\small (Color online). Schematic entanglement classification with marginal passive state energies. Here, the passive state means it has no extracted energy under any cyclic unitary process.}
\label{figure1}
\end{figure}

Multipartite quantum states play a crucial role in various quantum technologies such as quantum communication, quantum computing, and interferometry. An important question that arises in the study of these states is whether they can be classified based only on local information. The investigation into this marginal problem originated in the context of the well-known Pauli exclusion principle for fermions \cite{Coleman,Klyachko}. In general, the classification of entanglement involves examining how entanglement behaves under stochastic local operations and classical communication (SLOCC). Recent results have shown that for pure quantum states, information about individual particles alone can be used to detect multipartite entanglement \cite{Walter2013}. Specifically, the spectral vectors of reduced densities of individual subsystems give rise to an entanglement polytope corresponding to each entanglement class under SLOCC. Violations of these generalized polytope inequalities provide an effective method for detecting multipartite entanglement locally. Experimental validation of this entanglement classification approach has been achieved \cite{Aguilar2015,Zhao2017}. However, the direct connection between entanglement classification under SLOCC and passive state energy remains unclear. A schematic representation of entanglement classification with passive state energies is depicted in Fig. \ref{figure1}.

In this paper, we investigate the relationship between entanglement classification under SLOCC and passive state energy. We focus on the marginal passive state energy in the bipartition scenario involving a single qubit and the remaining qubits. We establish a link between the marginal passive state energies and the geometric entanglement measure. By using polygon inequalities, we prove for multi-qubit pure states that each marginal passive state energy does not exceed the sum of the others. Moreover, we extend to demonstrate that the vectors of marginal passive state energies collectively form a distinct polytope for each entanglement class. The present passive state energy polytope not only complements the entanglement polytope but also provides an additional criterion for discerning SLOCC multipartite entanglement classes. We further introduce an indicator related to average energies to distinguish between intertwined entanglement polytopes \cite{Walter2013}. One example is the generalized W states and GHZ states. This indicator highlights SLOCC entanglement classes that cannot be identified using the entanglement polytopes \cite{Walter2013}.

\section{The marginal passive qubit states}

Consider a finite-dimensional system described by the state $\varrho$ on Hilbert space $\mathcal{H}$ with Hamiltonian $H=\sum^d_{i=1}\epsilon_iE|\epsilon_i\rangle\langle\epsilon_i|$ with $\epsilon_1 \leq \cdots \leq \epsilon_{d}$ and unit energy $E$. The energy of the system is denoted as  $E(\varrho)={\rm Tr}(\varrho H)$. According to the Schrodinger dynamics, the overall unitary evolution operator $U(\tau)$ yields the final state to be $\varrho(\tau)=U(\tau)\varrho U^{\dagger}(\tau)$. A state is said to be passive if its average energy cannot be decreased through cyclic unitary evolutions  \cite{Allahverdyan2004}, that is
\begin{eqnarray}
{\rm Tr}(\varrho(\tau)H)\geq{\rm Tr}(\varrho H),
\end{eqnarray}
which means that no energy can be extracted from the state through a unitary process, as lowering the system's energy would entail transferring it to an energy storage device, in accordance with energy conservation principles.

In general, an $n$-qubit entangled pure state on Hilbert space $\mathcal{H}_1 \otimes \cdots \otimes \mathcal{H}_n$ can be written into
\begin{eqnarray}
|\Psi\rangle=\sum_{j=0}^n \sum_{s_j=0}^1 \alpha_{s_1 \cdots s_n}\left|s_1 \cdots s_n\right\rangle,
\end{eqnarray}
where $\alpha_{s_1 \cdots s_n}$ are the coefficients satisfying the normalization condition of $\sum_{s_1 \cdots s_n}\left|\alpha_{s_1 \cdots s_n}\right|^2=1$. Define the reduced density matrix of the qubit $i$  as $\varrho_i, i=1, \cdots, n$,  there are two eigenvalues $\{\lambda_{\min }^{(i)}, 1-\lambda_{\min }^{(i)}\}$ with $\lambda_{\min }^{(i)} \in[0, \frac{1}{2}]$.
Consider all the marginal passive state energies for an $n$-qubit state with subsystem $i$, denoted as $E_{i}$. Herein, each qubit $i$ is assumed to be
governed by the Hamiltonian $H_i=E|1\rangle\langle 1|$ under local unitary operations for $1 \leq i\leq n$. The composite system is governed by the Hamiltonian $H=\sum^n_{i=1}H_i\otimes_{k\in\overline{i}} {\mathbb I}_{k}$, herein, $H_i\otimes_{k\in\overline{i}} {\mathbb I}_{k}={\mathbb I}_{1}\cdots {\mathbb I}_{i-1}\otimes {H}_{i}\otimes{\mathbb I}_{i+1}\cdots {\mathbb I}_{n}$. The  energy  of the marginal passive state (EMPS) is then given by
\begin{eqnarray}
\mathcal{E}_{i}(|\Psi\rangle)=\lambda^{(i)}_{\min}E=\frac{{\cal G}(|\Psi\rangle_{i|\overline{i}})E}{2},
\label{puregap1}
\end{eqnarray}
where GME denotes the geometric measure of entanglement \cite{Wei2003}. This reveals a remarkable correspondence between the thermodynamic quantity and the operational information for any $2\times 2^{n-1}$-dimensional isolated system.

Consider the spectrum of the reduced density matrices  of any $n$-qubit pure state $|\Psi\rangle$ on Hilbert space $\mathcal{H}_{1}\otimes \cdots \otimes \mathcal{H}_n$, the following polygon inequalities hold\cite{Higuchi2003}:
\begin{eqnarray}
\lambda^{(i)}_{min}\leq \sum_{j\neq i}\lambda^{(j)}_{min},
\label{polygon}
\end{eqnarray}
where $\lambda^{(i)}_{min}\in[0,\frac{1}{2}]$ is the smallest eigenvalue of the reduced density matrix $\varrho_i$ of the $i$-th qubit. According to Eq.(\ref{puregap1}) and the inequality (\ref{polygon}), it implies  the following polygon inequalities for the EMPS:
\begin{eqnarray}
\mathcal{E}_{i}(|\Psi\rangle)\leq \sum_{j\neq i}\mathcal{E}_{j}(|\Psi\rangle)
\label{gappolygon}
\end{eqnarray}
with the local Hamiltonian $H_i=E|1\rangle\langle 1|$ for the $i$-th qubit,

\begin{figure}
\begin{center}
\resizebox{250pt}{110pt}{\includegraphics{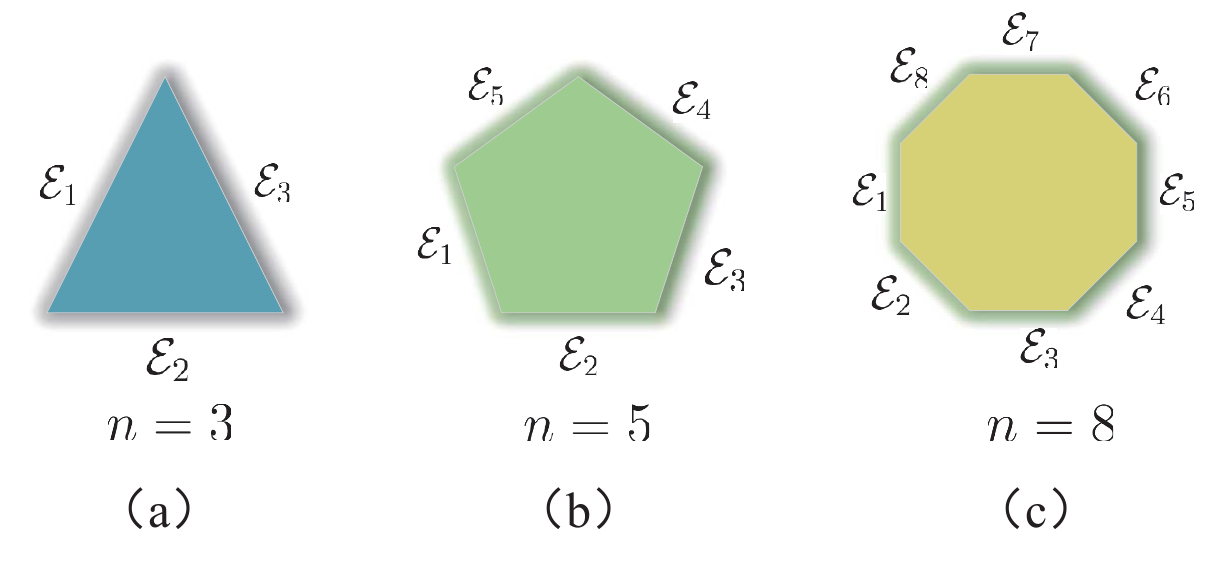}}
\end{center}
\caption{\small (Color online). Schematic polygon inequalities of EMPS. (a) 3-qubit pure states; (b) 5-qubit pure states; (c) 8-qubit pure states. The length of each side corresponds to the value of EMPS.}
\label{polygontu}
\end{figure}

Remarkably, the inequality (\ref{gappolygon}) guarantee that the EMPSs form a closed $n$-sided polygon. This allows for a geometric representation of the inequality (\ref{gappolygon}) as shown in Fig.\ref{polygontu}. The inequality (\ref{gappolygon}) first describes the polygon properties of thermodynamic quantities,  going beyond entanglement measure\cite{Qian2018,Luo2022}, and differing from monogamous characteristics\cite{Bai3}.

Let $\mathcal{E}^{\textsf{Tol}}(|\Psi\rangle)$ denote the total EMPS for all possible marginal subsystems, that is,
\begin{eqnarray}
\mathcal{E}^{\textsf{Tol}}(|\Psi\rangle)=\sum^n_{j=1}\mathcal{E}_{j}(|\Psi\rangle).
\label{energyTotal}
\end{eqnarray}
This thermodynamic quantity shows the following relation to the EMPS of single qubit as
\begin{eqnarray}
\mathcal{E}^{\textsf{Tol}}(|\Psi\rangle)\geq 2\mathcal{E}_{j}(|\Psi\rangle), j=1, \cdots, n.
\label{aa}
\end{eqnarray}

\section{Multi-qubit entanglement classification}

The entanglement among quantum systems is highly relevant in thermodynamics, as passive-state energy as a thermodynamic measure of entanglement \cite{Alimuddin2020}. Our goal here is to explore many-body entanglement classification using the EMPSs based on the quantum marginal problem related to the $N$-representability problem in quantum chemistry \cite{Walter2013}. Specially, consider an $n$-qubit quantum state $|\Psi\rangle$ on Hilbert space $\otimes_{i=1}^n\mathcal{H}_i$. It exhibits entanglement if it cannot be expressed as the product of individual qubit states~\cite{Horodecki2009}. Two states are deemed equivalent under SLOCC  if and only if they can be converted into one another through local invertible operations. Define a characteristic vector of the EMPS as
\begin{eqnarray}
\textbf{E}(|\Psi\rangle)=(\mathcal{E}_{1}(|\Psi\rangle),\mathcal{E}_{2}(|\Psi\rangle),\cdots,\mathcal{E}_{n}(|\Psi\rangle))
\label{puregapN}
\end{eqnarray}
where $\mathcal{E}_{i}(|\Psi\rangle)$ satisfies the inequality (\ref{gappolygon}). Denote ${\cal C}=G\cdot{}\rho$ as the SLOCC entanglement class containing $\rho$, where $G$ denotes a local invertible action such that $\varrho=G\cdot\rho$. From the inequality (\ref{gappolygon}) all the EMPSs of any state in $\mathcal{C}$ consist of a convex polytope.

\begin{theorem}
The EMPS polytope of an entanglement class $\mathcal{C}=G\cdot|\Psi\rangle$ for a given entangled pure state $|\Psi\rangle$ on Hilbert space $\otimes_{i=1}^n\mathcal{H}_i$ is given by the convex hull
\begin{eqnarray}
\Omega_{\cal C}={\rm conv}\{\rm{\textbf {E}}(|\Psi\rangle),\forall |\Psi\rangle\}.
\end{eqnarray}
\label{convex}
\end{theorem}

\emph{Proof}.
For an  $n$-qubit system. Define the marginal reduced density matrix of the qubit $i$ as $\varrho_i$. For each $\varrho_i$, let  $\lambda^i_{\max }$ be the largest  eigenvalue. The vectors $\vec{\lambda}=(\lambda^{(1)}_{\max}, \lambda^{(2)}_{\max},\cdots,\lambda^{(n)}_{\max})$
 corresponding to all pure states  within the closure of an orbit under SLOCC transformations constitute a polytope\cite{Walter2013}. Combining with \cite[Theorem 2]{Walter2013}, it follows Theorem \ref{convex} using the relation (\ref{puregap1}). This has completed the proof. $\Box$

Theorem \ref{convex} reveals thermodynamics  meanings of entanglement and provides a criterion for entanglement classification under SLOCC.

\begin{figure}
\begin{center}
\resizebox{250pt}{120pt}{\includegraphics{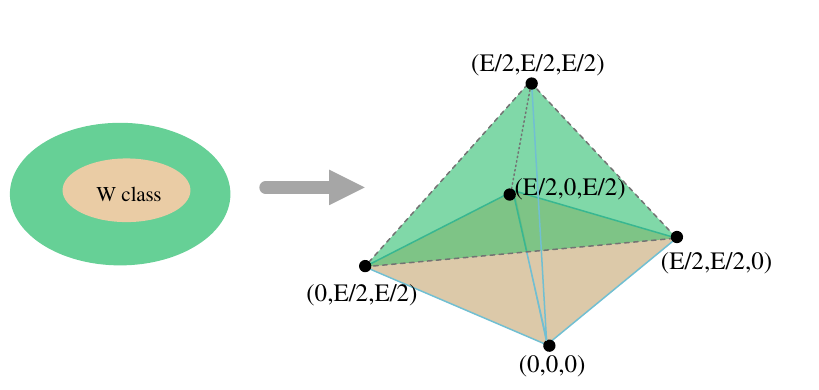}}
\end{center}
\caption{\small (Color online). The EMPS polytopes of three-qubit pure states.}
\label{polytope}
\end{figure}

For a three-qubit pure state, there exists a one-to-one correspondence between six entanglement classes and the EMPSs. Combining with the entanglement polytopes \cite{Walter2013}, Theorem \ref{convex} implies that the EMPS polytopes of all pure states consist of a convex hull with five vertices as shown in Fig.\ref{polytope}.
\begin{itemize}
\item   Fully separable states. This kind of states  corresponds to one vertex of $\Lambda_{\mathcal{S}}=(\mathcal{E}_{1},\mathcal{E}_{2},\mathcal{E}_{3})=(0,0,0)$;

  \item  Biseparable states (BS). In addition to rearranging the parties and applying local unitary transformations, the three classes of states can be exemplified by the state
$|\psi\rangle_{\mathcal{B S}}=|0\rangle(\alpha|00\rangle+\beta|11\rangle)$, where
$|\alpha|^2+|\beta|^2=1$. Focus on one of these cases, as the other two are obtained through simple label permutations. This defines the entanglement polytope, characterized by
$0 \leq \mathcal{E}_{2}+\mathcal{E}_{3}\leq 1$. These polytopes are depicted by the thick straight lines in Fig. \ref{polytope}, originating at the vertex (0,0,0). Three vertices, $(0,\frac{E}{2},\frac{E}{2})$, $(\frac{E}{2},0,\frac{E}{2})$ and $(\frac{E}{2},\frac{E}{2},0)$, are for three kinds of biseparable states $|BS\rangle=\frac{1}{\sqrt{2}}(|00\rangle+|11\rangle)\otimes|0\rangle$ and its permutations.

     \item  The W-class states (the orange tetrahedron).  These states can be represented by
     \begin{eqnarray}
|W\rangle=a_1|001\rangle+a_1|010\rangle+a_2|100\rangle
\label{Wstate}
\end{eqnarray}
 with $\sum_i|a_i|^2=1$.  For any W-type entanglement (\ref{Wstate}), the vector $(\lambda^{(1)}_{\max}, \lambda^{(2)}_{\max}, \lambda^{(3)}_{\max})$, representing the set of local maximal eigenvalues, resides within the W-type entanglement polytope \cite{Walter2013}. This implies that
\begin{eqnarray}
\lambda^{(1)}_{\max}+\lambda^{(2)}_{\max}+\lambda^{(3)}_{\max}\geq 2.
\label{GHZ1}
\end{eqnarray}
Assume that the local Hamiltonian
 $H=E|1\rangle\langle 1|$. Combining with the inequality (\ref{GHZ1}) and Eq.(\ref{puregap1}),  it follows that the EMPS vector satisfies the following relation as
\begin{eqnarray}
\mathcal{E}_1+\mathcal{E}_2+\mathcal{E}_3\leq E.
\label{GHZ}
\end{eqnarray}
  \item  The GHZ-class state (the entire polytope).
The entire polytope is for the generalized GHZ states:
\begin{eqnarray}
|GHZ\rangle=\cos\theta|000\rangle+\sin\theta|111\rangle, \theta\in (0, \frac{\pi}{4}).
\label{GHZstate}
\end{eqnarray}
\end{itemize}

The inequality (\ref{GHZ}) provides a new classification method for three-qubit pure states under the SLOCC, as shown in Fig.\ref{polytope}. Violating the inequality (\ref{GHZ}) by the EMPS witnesses the GHZ-type entanglement. Howbeit, an EMPS vector may fail to fully distinguish different entanglement classes. This can be further resolved by using the following energy indicator as
\begin{eqnarray}
\eta_{E}(|\Psi\rangle)=\min_{j} \left\{\sum^n_{k\neq j,k=1}{\mathcal{E}}_{k}(|\Psi\rangle)-\mathcal{E}_{j}(|\Psi\rangle)
\right\},
\label{indicatorME}
\end{eqnarray}
The quantity $\eta_{E}(|\Psi\rangle)$ characterizes the genuine tripartite entanglement, i.e., it cannot be decomposed into biseparable states in terms of any bipartition $S$ and $\overline{S}$ as $|\Psi\rangle=$ $|\psi\rangle_S \otimes|\phi\rangle_{\overline{S}}$. Otherwise, it is biseparable state.

\begin{prop}
\label{Threequbit}
Let each qubit be governed by the Hamiltonian $H_j=E|1\rangle\langle1|$. A three-qubit pure state $|\Psi\rangle$ is genuine tripartite entangled if $\eta_{E}(|\Psi\rangle)\neq 0$.
\end{prop}

\textit{Proof}. Assume that the quantum state $|\Psi\rangle$ on the Hilbert space $\mathcal{H}_1\otimes \mathcal{H}_2\otimes \mathcal{H}_3$ is not genuinely tripartite entangled, i.e., it can be decomposed as follows:
\begin{eqnarray}
&&|\Psi_{ijk}\rangle=|\varphi_{i}\rangle\otimes|\varphi_{jk}\rangle,
\label{bipartite0}
\\
&&|\Psi_{ijk}\rangle=|\phi_{i}\rangle\otimes|\phi_{j}\rangle\otimes|\phi_{k}\rangle,
\label{bipartite1}
\end{eqnarray}
where $i\neq j\neq k\in\{1,2,3\}$, $|\varphi_{i}\rangle$ and $|\phi_j\rangle$ represent the states of systems $i$ and $j$, respectively, and $|\varphi_{jk}\rangle$ denotes the state of the joint system $jk$. In the case of Eq.(\ref{bipartite0}), it follows from Eq.(\ref{puregap1}) that $\mathcal{E}_i=0$, and $\mathcal{E}_j=\mathcal{E}_k$ from the pure state $|\varphi_{jk}\rangle$. Moreover, we can show that $\mathcal{E}_j=\mathcal{E}_i+\mathcal{E}_k$ for all $i\neq j\neq k\in\{1,2,3\}$, where $\mathcal{E}_{j}$ represents the average extracted energy of the marginal passive state for the subsystem $j$. This implies $\eta_{E}(|\Psi\rangle)=0$ for any $j\in \{1,2,3\}$. A similar result holds for the case (\ref{bipartite1}). $\Box$

Proposition \ref{Threequbit}
provides a method to  verify the genuine tripartite entanglement. To show the main idea we consider the generalized GHZ states
$
\left|\psi_g\right\rangle=\cos \theta_1|000\rangle+\sin \theta_1|111\rangle,
$ where $\theta_1\in[0,\pi/4]$, each qubit is governed by the bare Hamiltonian $H=E|1\rangle\langle1|$.  It is easy to check that $\tau(|\psi_g\rangle)=\sin^22\theta_1\neq0$ if and only if $\theta_1\neq0$. Here, $\tau$ is a genuine tripartite entanglement measure \cite{Coffman2000}. On the other hand, we obtain that $\eta_{E}(|\psi_g\rangle)=\sin^2\theta_1E\neq0$ if and only if $\theta_1\neq0$. This implies that $|\psi_g\rangle$ is a genuine tripartite entanglement.

As a complement to Theorem 1, the indicator $\eta_{E}(|\Psi\rangle)$ provides a finer criterion for classifying multi-qubit pure states under the LOCC.

\section{Examples}

\textbf{Example 1}. Considering a generalized $n$-qubit W state on the Hilbert space $\otimes_{i=1}^n\mathcal{H}_i$ given by:
\begin{eqnarray}
|W_n\rangle=\sum^n_{i=1}\sqrt{a_i}|\vec{1}_i\rangle_{1\cdots n},
\label{Wn}
\end{eqnarray}
where $\vec{1}_i$ denotes all zeros except for the $i$-th component being $1$, and $a_i$ satisfies $\sum^n_{i=1}a_i=1$ with $a_1\geq\cdots \geq a_{n}$. Let's assume that each qubit $i$ is governed by the bare Hamiltonian $H_i=E|1\rangle\langle 1|$ for $1\leq i\leq n$. From Eq.(\ref{puregap1}), if there exists an $a_i>1/2$, it implies that $\mathcal{E}_{i}(|W_n\rangle)=(1-a_i)E$ and $\mathcal{E}_{j}(|W_n\rangle)=a_jE$ for $j\neq i$. Conversely, for all $a_i<\frac{1}{2}$, we obtain that  $\mathcal{E}_{i}(|W_n\rangle)=a_iE$ for any $1\leq i\leq n$. This further implies that
 \begin{eqnarray}
\mathcal {E}^{\textsf{Tol}}(|W_n\rangle)=
\left\{
\begin{aligned}
  &2(1-a_i)E<E,    && \exists i, a_i>\frac{1}{2},
  \\
  &E,    && \forall i, a_i<\frac{1}{2}.
\end{aligned}
\label{Wn0}
\right.
\end{eqnarray}
Hence, we obtain that $\mathcal{E}^{\textsf{Tol}}(|W_n\rangle)\leq E$ for any $n$-qubit W state. This means that the generalized W state with $a_i<1/2$ for any $i$ lies on the facet of $\mathcal{E}^{\textsf{Tol}}(|W_n\rangle)=E$.

\textbf{Example 2}. Consider a generalized $n$-qubit GHZ state on Hilbert space $\otimes_{i=1}^n\mathcal{H}_i$ given by
\begin{eqnarray}
|GHZ_n\rangle&=(\cos\theta|0\rangle^{\otimes n}+\sin\theta|1\rangle^{\otimes n})_{1\cdots n}
\label{GHZg}
\end{eqnarray}
with $\theta\in(0,\frac{\pi}{4}]$. For each qubit the bare Hamiltonian is given by $H=E|1\rangle\langle 1|$. According to Eq.(\ref{puregap1}), the EMPS for the subsystem $i$ is given by $\mathcal{E}_{i}(|GHZ_g\rangle)=\sin^2\theta E$. This implies that
\begin{eqnarray}
\mathcal{E}^{\textsf{Tol}}(|GHZ_n\rangle)=n\sin^2\theta E\leq nE,
\label{GHZn1}
\end{eqnarray}
where the maximum EMPS is obtained from the maximally entangled $n$-qubit GHZ state, i.e.,  $\theta=\pi/4$. Moreover, we have
\begin{eqnarray}
\mathcal{E}^{\textsf{Tol}}(|GHZ_n\rangle)>E
\label{GHZn2}
\end{eqnarray}
for $\theta>\arcsin \sqrt{1/n}$. The generalized GHZ state with $\theta>\arcsin\sqrt{1/n}$ does not belong to the facet determined by the generalized $n$-qubit W state in Example 1. This provides a way to distinguish two entanglement classes by using the facet of $\mathcal{E}^{\textsf{Tol}}=E$.

For the case of $\theta<\arcsin \sqrt{1/n}$, we obtain that
\begin{eqnarray}
\mathcal{E}^{\textsf{Tol}}(|GHZ_n\rangle)<E.
\label{GHZn}
\end{eqnarray}
According to Eqs. (\ref{GHZn}) and (\ref{Wn0}), the presence of overlapping regions fails to distinguish W-type entanglement from GHZ-type entanglement. The indicator $\eta_{E}$ in Eq.(\ref{indicatorME}) may be used to certify the entanglement as
\begin{eqnarray}
&&\eta_{E}(|W_n\rangle)=0,\exists i, a_i>\frac{1}{2},
\\
&&\eta_{E}(|GHZ_n\rangle)=(n-2)\sin^2\theta E>0,~~ \forall i.
\end{eqnarray}
Moreover, for the case of $\theta=\arcsin \sqrt{1/n}$, we obtain
\begin{eqnarray}
\eta_{E}(|GHZ_n\rangle)=\frac{(n-2)E}{n}.
\label{GHs}
\end{eqnarray}
Meanwhile, for the generalized W state we obtain
\begin{eqnarray}
&&\eta_{E}(|W_n\rangle)=(1-2a_i)E,\forall i, a_i<\frac{1}{2}.
\end{eqnarray}
This implies $\eta_{i}(|GHZ_n\rangle)\not=\eta_{i}(|W_n\rangle)$ for any one $a_i$ satisfying $a_i\not=1/n$. This means the generalised GHZ state can be distinguished from all the almost generalized W states except for the maximally entangled W state beyond recent result \cite{Walter2013}.

\textbf{Example 3}. Consider an $n$-qubit Dicke state with $l$ excitations \cite{Stockton2003} :
\begin{eqnarray}
|D^{(l)}_n\rangle_{12\cdots n}=\frac{1}{\sqrt{\binom{n}{l}}}\sum_{\textsf{g}\in S_n}\textsf{g}(|0\rangle^{\otimes n-l}|1\rangle^{\otimes l}),
\end{eqnarray}
where $1\leq l\leq n-1$. Here, the summation is taken over every possible permutation on $\textsf{g}\in S_n$ of the product states containing $n-l$ number of $|0\rangle$ and $l$ number of $|1\rangle$, $S_n$ denotes the permutation group on $n$ items, and $\binom{n}{l}$ denotes the combination number of choosing $l$ items from $n$ items. For each qubit the bare Hamiltonian is given by $H=E|1\rangle\langle 1|$. Due to the symmetry of the $n$-qubit symmetric Dicke states, the reduced density matrix of any subsystem $i$ is:
\begin{eqnarray}
\varrho_i=\frac{n-l}{n}|0\rangle\langle0|+\frac{l}{n}|1\rangle\langle1|.
\end{eqnarray}
Using Eq.(\ref{puregap1}), we find:
\begin{eqnarray}
\mathcal{E}_i(|D^{(l)}_n\rangle)=\frac{lE}{n} \quad \text{for } l\leq\frac{n}{2},
\end{eqnarray}
and for $l>\frac{n}{2}$:
\begin{eqnarray}
\mathcal{E}_i(|D^{(l)}_n\rangle)=\frac{nE-lE}{n}.
\end{eqnarray}
According to Eq.(\ref{energyTotal}), the total energy follows a general polytope facet:
\begin{eqnarray}
\mathcal{E}^{\textsf{Tol}}(|D^{(l)}_n\rangle)=
\left\{
\begin{aligned}
&  lE,    & \text{if } l\leq\frac{n}{2};
  \\
&  nE-lE,    & \text{if } l>\frac{n}{2}.
\end{aligned}
\right.
\label{DickeN}
\end{eqnarray}
In particular, when $l=1$ and $n\geq3$, the $n$-qubit Dicke state reduces to the $n$-qubit W state, leading to $\mathcal{E}^{\textsf{Tol}}(|W\rangle)=E$. For any generalized Dicke states:
\begin{eqnarray}
|\hat{D}^{(l)}_n\rangle=\sum_{\textsf{g}\in S_n}\alpha_{\textsf{g}}\textsf{g}(|0\rangle^{\otimes n-l}|1\rangle^{\otimes l})_{12\cdots n},
\end{eqnarray}
where the coefficients $\alpha_{\textsf{g}}$ depend on the permutation $\textsf{g}\in S_n$ and satisfy $\sum^{\binom{n}{l}}_{\textsf{g}}\alpha_{\textsf{g}}^2=1$. Due to symmetry, $\mathcal{E}_i(|\hat{D}^{(l)}_n\rangle)$ is maximized when $\alpha_{\textsf{g}}=1/\sqrt{\binom{n}{l}}$ for each $\alpha_{\textsf{g}}$, leading to:
\begin{eqnarray}
\mathcal{E}^{\textsf{Tol}}(|\hat{D}^{(l)}_n\rangle)\leq
\left\{
\begin{aligned}
&  lE,    & \text{if } l\leq\frac{n}{2}; \\
 & nE-lE,    & \text{if } l>\frac{n}{2}.
\end{aligned}
\right.
\end{eqnarray}
This shows that generalized Dicke states are below the facet of $\mathcal{E}^{\textsf{Tol}}(|\hat{D}^{(l)}_n\rangle)< lE$, while $n$-qubit Dicke states $|D^{(l)}_n\rangle$ lie on the facet of $\mathcal{E}^{\textsf{Tol}}(|D^{(l)}_n\rangle)=lE$. This observation provides a method for distinguishing Dicke states with different excitations that are inequivalent under the SLOCC, indicating that $n$-qubit Dicke states with different excitations lie on different facets.

\textbf{Example 4}. Consider a noisy W state as
\begin{eqnarray}
\varrho_{W}=(1-v_1)|W\rangle\langle W|+\frac{v_1}{8}\mathbbm{1}_8,
\label{Wmixed}
\end{eqnarray}
and the noisy GHZ state
\begin{eqnarray}
\varrho_{GHZ}=(1-v_2)|GHZ\rangle\langle GHZ|+\frac{v_2}{8}\mathbbm{1}_8,
\label{GHZmixed}
\end{eqnarray}
where $\mathbbm{1}_8$ denotes the identity matrix of rank $8$, $|W\rangle$ and $|GHZ\rangle$ are maximally entangled states which are respectively defined as $|W\rangle=\frac{1}{\sqrt{3}}(|100\rangle+|010\rangle+|001\rangle)$ and  $|GHZ\rangle=\frac{1}{\sqrt{2}}(|000\rangle+|111\rangle)$. Here, each qubit $i$ is governed by the Hamiltonian $H_i=E|1\rangle\langle 1|$ for $1\leq i\leq 3$. These states are genuinely multipartite entangled for $v_1<9/17$ \cite{separable2010} and $v_2<4/7$ \cite{Huber2010}, respectively.

Consider any subsystem $i\in\{A,B,C\}$, the spectrum of reduced state $\varrho_i$ for $\varrho_{W}$ are given by $\{(4-v_1)/6,(2+v_1)/6\}$. The extremal energy of the subsystem $i\in\{A,B,C\}$ is shown as $\mathcal{E}_i(\varrho_{W})=(2E+v_1E)/6$. From Eq.(\ref{energyTotal}) we obtain that
\begin{eqnarray}
\mathcal{E}^{\textsf{Tol}}(\varrho_{W})=\sum_i\mathcal{E}_i(\varrho_{W})=\frac{2E+v_1E}{2}<\frac{43E}{34}
\end{eqnarray}
for $0<v_1<9/17$. For noisy GHZ states, the reduced states of single qubit are completely mixed state $\frac{1}{2}\mathbbm {I}$. This implies that  $\mathcal{E}_i(\varrho_{GHZ})=E/2$, which further follows that
\begin{eqnarray}
\mathcal{E}^{\textsf{Tol}}(\varrho_{GHZ})=\frac{3E}{2}>\mathcal{E}^{\textsf{Tol}}(\varrho_{W}).
\end{eqnarray}
So, the quantity of $\mathcal{E}^{\textsf{Tol}}$ can be used to distinguish the noisy W states from the noisy GHZ states.

Using the average extracted energy of the passive state for the subsystem allows distinguishing the W-type entanglement from the GHZ-type entanglement. This is going beyond the capabilities of entanglement polytopes \cite{Walter2013}. The following example shows that the energy indicator can be applied to characterize the  genuine multipartite entanglement.

\begin{figure}
\begin{center}
\resizebox{250pt}{120pt}{\includegraphics{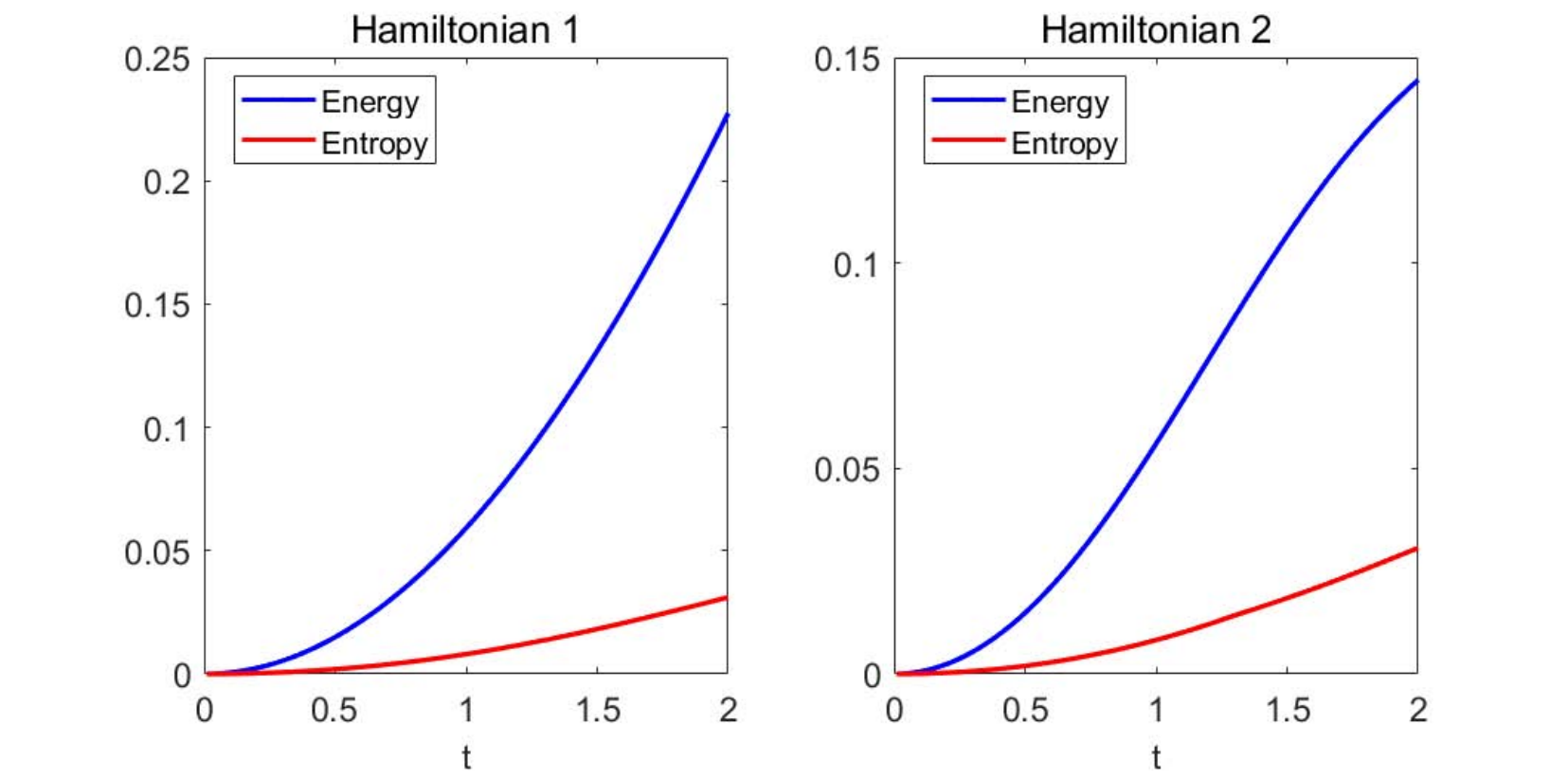}}
\end{center}
\caption{\small (Color online). Genuinely multipartite entanglement of Ising chain model. In simulation, the  Hamiltonian 1 is defined in Eq.(\ref{H1}) with $N= 5$,  the coupling constant $J =1$ and external magnetic field $h =1$. The Hamiltonian 2 is defined in Eq.(\ref{H2}) with long-range correlations. Energy ($\eta_E/E$) denotes the energy indicator of the marginal passive states with bare Hamiltonian $H=E|1\rangle \langle 1|$. Entropy denotes the entanglement measure $\min_{i,j}|S(\rho_{ij})-S(\rho_i)-S(\rho_j)|$, where $S$ denotes the von Neumann entropy and $\rho_i,\rho_{ij}$  denote the reduced matrices.
}
\label{Simulation}
\end{figure}

\textbf{Example 5} Consider a chain of $N$ spins, where each spin can be either up (+1) or down (-1). The Hamiltonian is given by
\begin{eqnarray}
H_0 = -J \sum_{i=1}^{N-1} s_i s_{i+1} - h \sum_{i=1}^{N} s_i,
\end{eqnarray}
Here, $J$ represents the coupling constant that signifies the intensity of interaction among adjacent spins, $h$ is the external magnetic field, and $s_i$ represents the spin at site $i$. In this model, the spins tend to align with their neighbors to minimize the energy of the system. This model can be rewritten in terms of Pauli matrices into
\begin{eqnarray}
H = -J \sum_{i=1}^{N-1} s_i^z s_{i+1}^z - h \sum_{i=1}^{N} s_i^z,
\label{H1}
\end{eqnarray}
where $s_z = \frac{1}{2} \sigma_z$. The simulation of $\eta$ defined in Eq.(\ref{indicatorME}) is shown in Fig.\ref{Simulation}. Here, the genuine multipartite entanglement can be verified by using a simple standard from the additive of von Neumann entropy, i.e., it is genuinely multipartite entangled pure state if $\min_{ij}|S(\rho_{ij})-S(\rho_i)-S(\rho_j)|>0$, where $S$ denotes the von Neumann entropy and $\rho_i,\rho_{ij}$  denote the reduced matrices. This kind of the nearest-correlated system can generate the genuine multipartite entanglement. Moreover, we consider another system with long-range correlations. The Hamiltonian is given by
\begin{eqnarray}    H=H_0+4\sigma^x_2\sigma^x_3\sigma^x_4+3\sigma^x_1\sigma^x_3\sigma^x_4\sigma^x_5+3\sigma^x_1\sigma^x_2\sigma^x_4\sigma^x_5.
\label{H2}
\end{eqnarray}
Both entanglement measure can be used to characterize the genuine multipartite entanglement. Although it is difficult to characterize all the genuine multipartite entanglement, the present energy indicator is useful for pure states. It is valuable for further exploration of entanglement classification and entanglement measure.

\textbf{Conclusion}. We have established a correspondence between the geometric measure of entanglement and EMPS. We proved a polygon inequality for arbitrary $n$-qubit pure states in terms of the EMPS. Associated with these inequalities, we have shown that the possible sets of all marginal passive state energies form a convex polytope. This reveals a physical interpretation of the entanglement polytopes, as well as provides a tool to detect entanglement in multi-qubit systems. Moreover,  a set of multipartite EMPS indicators has been introduced to distinguish the overlapping polytope of different entanglements in ref.\cite{Walter2013}. This provides new insights of entanglement classifications going beyond the one-body quantum marginal quantities.

Given the profound analogies between thermodynamic quantities and entanglement, the passive-state energy provides a good entanglement measure for the pure bipartite states \cite{Alimuddin2020}. This may lead to the possibility of capturing the signature of genuineness in multipartite entanglement from the passive state energy. Furthermore, it is still unknown whether the set of EMPS for multipartite mixed states forms a convex polytope.

\textbf{Acknowledgments}. This work was supported by the National Natural Science Foundation of China (Nos.62172341, 12204386, 12075159, and 12171044), Sichuan Natural Science Foundation (Nos.2023NSFSC0447, 24NSFSC5421), Beijing Natural Science Foundation (No.Z190005), Interdisciplinary Research of Southwest Jiaotong University China Interdisciplinary Research of Southwest Jiaotong University China (No.2682022KJ004), and the Academician Innovation Platform of Hainan Province.

\section*{Conflict of Interest}

The authors declare no other conflict of interest.

\section*{Author Contributions}

X.Y. and M.X.L. conducted the research. All authors wrote and reviewed the manuscript.

\section*{Data Availability Statement}

This is no data generated in research.




\begin{thebibliography}{00}

\bibitem{Allahverdyan2004}A. E. Allahverdyan, R. Balian, and Th. M. Nieuwenhuizen, Maximal work extraction from finite quantum systems, Europhys. Lett. 67, 565 (2004).

\bibitem{Viguie2005}  V. Viguie, K. Maruyama, and V. Vedral, Work extraction from tripartite entanglement, New J. Phys. 7, 195 (2005).

\bibitem{Skrzypczyk2014}P. Skrzypczyk, A. Short, and S. Popescu, Work extraction and thermodynamics for individual quantum systems, Nat. Commun. 5, 4185 (2014).

\bibitem{Perarnau2015} M. Perarnau-Llobet, K. V. Hovhannisyan, M. Huber, P. Skrzypczyk, N. Brunner, and A. Ac\'{\i}n, Extractable work from correlations, Phys. Rev. X 5, 041011 (2015).

\bibitem{Vin(2016)} S. Vinjanampathy, and J. Anders, Quantum thermodynamics,rs, Contemp. Phys. 57, 545 (2016).

\bibitem{Ciampini2017} M. A. Ciampini,  L. Mancino,  A. Orieux,  C. Vigliar,  P. Mataloni,  M. Paternostro, and  M. Barbieri, Experimental extractable work-based multipartite separability criteria, npj Quant. Inf. 3, 10 (2017).

\bibitem{Francica2017} G. Francica, J. Goold,  F. Plastina, and  M. Paternostro,  Daemonic ergotropy: Enhanced work extraction from quantum correlations, npj Quant. Inf. 3, 12 (2017).

\bibitem{Andolina2019} G. M. Andolina, M. Keck, A. Mari, M. Campisi, V. Giovannetti, and M. Polini, Extractable work, the role of correlations, and asymptotic freedom in quantum batteries, Phys. Rev. Lett. 122, 047702 (2019).

\bibitem{Monsel2020} J. Monsel, M. Fellous-Asiani, B. Huard, and A. Auff\`{e}ves, The energetic cost of work extraction, Phys. Rev. Lett. 124, 130601 (2020).


\bibitem{Opatrny2021} T. Opatrny, A. Misra, and G. Kurizki, Work generation from thermal noise by quantum phase-sensitive observation, Phys. Rev. Lett. 127, 040602 (2021).

\bibitem{Pusz1978}W. Pusz and S. L. Woronowicz, Passive states and KMS states for general quantum systems, Commun. Math. Phys. 58, 273(1978).

\bibitem{Salvia2020} R. Salvia and V. Giovannetti, Energy upper bound for structurally stable-passive N-passive states, Quantum 4, 274 (2020)

\bibitem{Sparaciari(2017)}C. Sparaciari, D. Jennings, and J. Oppenheim, Energetic instability of passive states in thermodynamics, Nat. Commun. 8, 1895 (2017).

\bibitem{Perarnau2015a}M. Perarnau-Llobet, K. V. Hovhannisyan, M. Huber, P. Skrzypczyk, J. Tura, and A. Acin, Most energetic passive states, Phys. Rev. E 92, 042147(2015).

\bibitem{Alhambra2019} A. M. Alhambra, G. Styliaris, N. A. Rodriguez-Briones, J. Sikora, and E. Martin-Martinez, Fundamental limitations to local energy extraction in quantum systems, Phys. Rev. Lett. 123, 190601 (2019).

\bibitem{Bera2019}  M. N. Bera, A. Riera, M. Lewenstein, Z. B. Khanian, and A. Winter, Thermodynamics as a consequence of information conservation, Quantum 3, 121 (2019).

\bibitem{Uzdin2018}  R. Uzdin and S. Rahav, Global Passivity in Microscopic Thermodynamics, Phys. Rev. X 8, 021064 (2018).

\bibitem{Brown2016} E. G. Brown, N. Friis, and M. Huber,Passivity and practical work extraction using Gaussian operations, New J. Phys. 18, 113028(2016).

\bibitem{Alimuddin2020} M. Alimuddin, T. Guha, and P. Parashar, Independence of work and entropy for equal-energetic finite quantum systems: Passive-state energy as an entanglement quantifier, Phys. Rev. E 102, 012145 (2020).

\bibitem{Koukoulekidis2021} N. Koukoulekidis, R. Alexander, T. Hebdige, and D. Jennings,The geometry of passivity for quantum systems and a novel elementary derivation of the Gibbs state, Quantum 5, 411 (2021).

\bibitem{Yang2023} X. Yang, Y.-H. Yang, M. Alimuddin, R. Salvia, S.-M. Fei, L.-M. Zhao, S. Nimmrichter, and M.-X. Luo, Battery Capacity of Energy-Storing Quantum Systems, Phys. Rev. Lett. 131, 030402 (2023).

\bibitem{Coleman} A. J. Coleman and V. I. Yukalov, Reduced density matrices: coulson's challenge (Springer, 2000).

\bibitem{Klyachko}A. Klyachko, Quantum marginal problem and $N$-representability, J. Phys.: Conf. Ser. 36, 72 (2006).

\bibitem{Walter2013}M. Walter,  B. Doran,  D. Gross, and M. Christandl, Entanglement polytopes: multiparticle entanglement from single-particle information, Science 340, 1205 (2013).

\bibitem{Aguilar2015}G. H. Aguilar, S. P.  Walborn,  P. S. Ribeiro, and L. C.  C\'{e}leri, Experimental determination of multipartite entanglement with incomplete information, Phys. Rev. X 5, 031042 (2015).

\bibitem{Zhao2017} Y. Y. Zhao,   M. Grassl,  B. Zeng,  G. Y. Xiang,  C. Zhang,  C. F. Li, and  G. C.  Guo, Experimental detection of entanglement polytopes via local filters, npj Quant. Inf.  3, 1-7 (2017).

\bibitem{Wei2003}  T. C. Wei,  P. M. Goldbart, Geometric measure of entanglement and applications to bipartite and multipartite quantum states, Phys. Rev. A 68, 042307 (2003).

\bibitem{Higuchi2003}   A. Higuchi, A. Sudbery, and J. Szulc, One-qubit reduced states of a pure many-qubit state: polygon inequalities, Phys. Rev. Lett. 90, 107902 (2003).

\bibitem{Qian2018}  X. F. Qian, M. A. Alonso, and J. H. Eberly, Entanglement polygon inequality in qubit systems, New J. Phys. 20, 063012 (2018).

\bibitem{Luo2022} X. Yang, Y. H. Yang, and M. X. Luo, Entanglement polygon inequality in qudit systems, Phys. Rev. A  105, 062402 (2022).

\bibitem{Bai3} Y. K. Bai, Y. F. Xu, and Z. D. Wang, General monogamy relation for the entanglement of formation in multiqubit systems, Phys. Rev. Lett. 113, 100503 (2014).

\bibitem{Coffman2000} V. Coffman, J.  Kundu, W. K. Wootters, Distributed entanglement, Phys. Rev. A 61, 052306(2000).

\bibitem{Horodecki2009}R. Horodecki, P. Horodecki, M. Horodecki, and K. Horodecki, Quantum entanglement, Rev. Mod. Phys. 81, 865 (2009).

\bibitem{Stockton2003}    J. K.  Stockton,  J. M. Geremia,  A. C. Doherty, and  H.  Mabuchi, Characterizing the entanglement of symmetric many-particle spin-1/2 systems. Phys. Rev. A  67, 426-430 (2003).

\bibitem{separable2010} O. G\"{u}hne and M. Seevinck, Separability criteria for genuine multiparticle entanglement, New J. Phys. 12, 053002(2010).

\bibitem{Huber2010}  M. Huber,  F. Mintert,  A, Gabriel, and B. C. Hiesmayr, Detection of high-dimensional genuine multipartite entanglement of mixed states, Phys. Rev. Lett.  104, 210501(2010).

\end{thebibliography}
\end{document}